%% file: ms.tex
\documentclass[twocolumn, superscriptaddress]{revtex4-1} \usepackage{amsmath} \usepackage{booktabs}
\usepackage[utf8]{inputenc} \usepackage{xcolor} \usepackage{graphicx} \usepackage{hyperref}
\usepackage{orcidlink}
\usepackage{lineno}

\renewcommand{\eqref}[1]{Eq.~(\ref{#1})}

\newcommand{\figref}[1]{Fig.~\ref{#1}}
\newcommand{\secref}[1]{Sec.~\ref{#1}}
\newcommand{\appref}[1]{Appendix~\ref{#1}}

\begin{document}
\title{Background Filter: A method for removing signal contamination during significance estimation of a GstLAL anaysis}

\input{author_list}
\date{\today}

\begin{abstract}
To evaluate the probability of a gravitational-wave candidate originating
from noise, GstLAL collects noise statistics from the data it analyzes. Gravitational-wave
signals of astrophysical origin get added to the noise statistics, harming the sensitivity
of the search. We present the Background Filter, a novel tool to prevent this
by removing noise statistics that were
collected from gravitational-wave candidates. To demonstrate its efficacy,
we analyze one week of LIGO and Virgo O3 data, and show that it improves the sensitivity 
of the analysis by 20-40\% in the high mass region, in the 
presence of 868 simulated gravitational-wave signals.
With the upcoming fourth observing run of LIGO, Virgo, and KAGRA expected to
yield a high rate of gravitational-wave detections,
we expect the Background Filter to be an important tool for
increasing the sensitivity of a GstLAL analysis.
\end{abstract}

\maketitle

\section{Introduction}
The Laser Interferometer Gravitational-wave Observatory (LIGO)~\cite{ligo} and
Virgo~\cite{virgo} collaborations have revolutionized
the field of gravitational-wave (GW) astronomy by detecting black hole and neutron
star mergers~\cite{LIGOScientific:2018mvr, gwtc-2, gwtc-2.1, LIGOScientific:2021djp}.
The detections have allowed us to observe
the universe in new ways and have opened up new avenues of scientific inquiry.~\cite{Abbott_2021, TGR, lensing, subsolarmass}
The GstLAL GW search pipeline~\cite{Messick:2016aqy, Sachdev:2019vvd, Hanna:2019ezx, softwarex}
(referred to as GstLAL hereafter) has been a
significant contributor to this field. In particular, GstLAL's ability to detect
signals in low-latency~\cite{170817_gcn} has facilitated multi-messenger
observations~\cite{170817_mm}.

GstLAL is a GW search pipeline that can process data from ground-based
GW detectors, such as the Hanford and Livingston LIGO detectors, the Virgo
detector and the KAGRA detector~\cite{kagra}, in near real time. It makes use of time-domain
matched-filtering to enable the detection of signals in noise-dominated data. It uses
a likelihood ratio (LR)~\cite{Tsukada:2023,Cannon:2015gha, Cannon:2012zt} as a ranking statistic for assigning
significance to detections. GstLAL divides its template bank~\cite{Mukherjee:2018yra, shio_template_bank}
into different ``template bins" to reduce the computational cost of the analysis,
and analyzes each one separately.
Some of these techniques are also used by other search
pipelines, such as PyCBC~\cite{Dal_Canton_2021, Davies_2020, pycbc}, MBTA~\cite{mbta, Adams_2016},
SPIIR~\cite{spiir, spiir_2017}, and IAS~\cite{ias, Zackay_2021}.

The fourth observing run of the LIGO Scientific, Virgo and KAGRA collaboration (O4) is set to
begin in May 2023~\cite{o4date} and promises to provide improved detector sensitivity.
GstLAL will continue to play an essential role in the detection of new GW
candidates. As such, it is necessary to keep refining the analysis pipeline to
reap the benefits of improved detector sensitivity to detect even more, and new
types of candidates. The Background Filter is one such new feature to this end.

This paper is structured as follows. In \secref{sec:signal_contamination},
we introduce the LR used by GstLAL, in particular the $\rho - \xi^2$ histograms
that GstLAL uses to evaluate one term of the likelihood ratio, and how the 
presence of GW signals in the data can cause ``contamination"
of these histograms. In \secref{sec:removing_contamination}, we describe how the
Background Filter works, and how it removes this contamination. Finally, in
\secref{sec:results}, we describe the analyses we performed to evaluate the
performance of the Background Filter, and the impact it has on the sensitivity of
a GstLAL analysis.

\section{Signal Contamination}
\label{sec:signal_contamination}

\subsection{Likelihood Ratio}

GstLAL is a matched-filtering based GW search pipeline which uses
a likelihood ratio statistic to rank GW candidates
~\cite{Tsukada:2023,Cannon:2015gha}. The LR is defined as
\begin{align}
    \mathcal{L}=\frac{P\left(\vec{O}, \vec{\rho}, \vec{\xi^2}, \vec{t}, \vec{\phi}, \bar{\theta} \mid \mathcal{H}_\mathrm{s} \right)}{P\left(\vec{O}, \vec{\rho}, \vec{\xi^2}, \vec{t}, \vec{\phi}, \bar{\theta} \mid \mathcal{H}_\mathrm{n} \right)},
\end{align}
where the numerator is the probability of obtaining a GW candidate
with parameters $(\vec{O}, \vec{\rho}, \vec{\xi^2}, \vec{t}, \vec{\phi}, \bar{\theta})$,
under the signal hypothesis ($\mathcal{H}_\mathrm{s}$) and the denominator is the
probability of obtaining the same candidate under the noise hypothesis ($\mathcal{H}_\mathrm{n}$).
$\vec{O}$ is the subset of GW detectors that the candidate was found in,
$\vec{\rho}$ is the set of matched-filtering signal-to-noise-ratios (SNRs) of those detectors,
$\vec{\xi^2}$ is the set of $\xi^2$-signal-based-veto values, $\vec{t}, \vec{\phi}$ are the
times and phases with which the candidate was found in the detectors, and $\bar{\theta}$
is the template which recovered the candidate, which also represents a set of 
intrinsic parameters (masses and spins).

The LR can be factorized as
\begin{widetext}
\begin{align}
\label{eq:full_lr}
    \mathcal{L}= \frac{P\left(\bar{\theta} \mid \mathcal{H}_\mathrm{s}\right)
        \times P\left(t_\mathrm{ref}, \phi_\mathrm{ref} \mid \bar{\theta}, \mathcal{H}_\mathrm{s}\right)
        \times P\left(\vec{O} \mid t_\mathrm{ref}, \mathcal{H}_\mathrm{s}\right)
        \times P\left(\vec{\rho}, \vec{\Delta t}, \vec{\Delta\phi} \mid \vec{O}, t_\mathrm{ref}, \mathcal{H}_\mathrm{s}\right)
        \times P\left(\vec{\xi^2} \mid \vec{\rho}, \bar{\theta}, \mathcal{H}_\mathrm{s}\right)
            }
        {P\left(t_\mathrm{ref}, \bar{\theta} \mid \mathcal{H}_\mathrm{n}\right)
        \times P\left(\vec{O} \mid t_\mathrm{ref}, \bar{\theta}, \mathcal{H}_\mathrm{n}\right)
        \times P\left(\vec{\Delta t}, \vec{\phi} \mid \vec{O}, \mathcal{H}_\mathrm{n}\right)
        \times P\left(\vec{\rho}, \vec{\xi^2} \mid t_\mathrm{ref}, \bar{\theta}, \mathcal{H}_\mathrm{n}\right)
            }
\end{align}
\end{widetext}

For a comprehensive explanation of \eqref{eq:full_lr} and every individual term in the LR,
readers are referred to~\cite{Tsukada:2023}. For the 
purpose of this paper, we are only concerned with the last term in the denominator,
$P\left(\vec{\rho}, \vec{\xi^2} \mid t_\mathrm{ref}, \bar{\theta}, \mathcal{H}_\mathrm{n}\right)$
(hereby referred to as the $\rho - \xi^2$ noise LR term).

\subsection{The $\rho - \xi^2$ histograms}

The $\rho - \xi^2$ noise LR term is calculated in a data-driven way.
GstLAL creates a histogram for each detector and template bin in $\rho - \xi^2$
space, called $\rho - \xi^2$ background histograms,
and populates it with the ($\rho$, $\xi^2$) values of noise events found in that
template bin during the analysis. Then, the $\rho - \xi^2$ noise LR term can be
calculated by evaluating the probability density function represented by the
histograms at the relevant ($\vec{\rho}$, $\vec{\xi^2}$) value.

Since the $\rho - \xi^2$ noise LR term assumes the noise hypothesis, we need to
populate the histograms with events originating only from noise, as compared to
events originating from GW candidates. To a large degree,
this is achieved by requiring those events to be
recovered only in one detector (called a single-detector or single event in contrast to a
coincident event) during a time when more than one detector
was producing data (called coincident time in contrast to single time).
This is because we expect GW signals to be correlated
across detectors, but not noise events.

\begin{figure*}
\includegraphics[width=\textwidth]{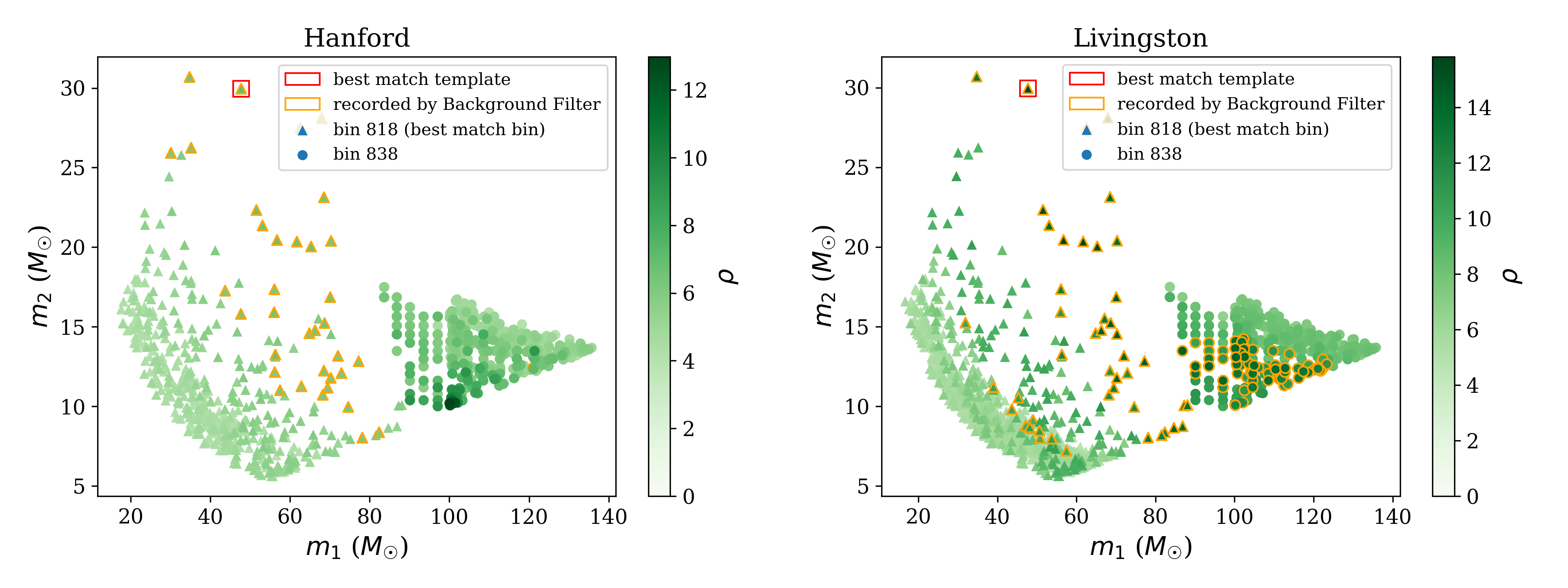}
\caption{\label{fig:svd}
An example of an event (GW200129\_065458) having templates with high match in
multiple template bins. Bin 818 has the best match with the GW
candidate, and recovers it in both Hanford and Livingston as a coincidence. Bin
838 has a lower match than bin 818, causing it to recover the candidate as a 
Livingston single. This will lead to the candidate being added to the $\rho - \xi^2$
background histogram of bin 838, causing signal contamination for bin 838. This
is shown in \figref{fig:signal_contamination}. The events passing the $\rho$ and
$\xi^2$ constraints, and hence recorded by the Background Filter are outlined in
orange.
}
\end{figure*}

Despite this, GW signals can sometimes enter the $\rho - \xi^2$
histograms. The reason might be astrophysical in origin, e.g. the
GW source is located in the blind spot of all
but one detector, or it might be terrestrial, e.g. only one detector is sensitive
enough to pick up the GW signal. In addition, GW
signals which are recovered as coincident events in one template bin
sometimes also get recovered as a single event, with a lower $\rho$ and LR
in other neighbouring template bins, which don't contain templates with high $\rho$
for that GW signal. We say a bin has a good match with a GW signal if it has
templates with a high $\rho$ for that GW signal, and that it has a bad match otherwise. 
GW events being recovered as coincident events in one bin and as single events in
others is demonstrated in \figref{fig:svd} for GW200129\_065458, a known
GW candidate reported in GWTC-3~\cite{LIGOScientific:2021djp}. The candidate is
recovered as a coincident event in bin 818, with which it has the best match.
It is also recovered in bin 838 as a Livingston single with a lower
$\rho$, since it's match with that bin isn't as good. As a result, the
candidate will be added to the background histogram of bin 838. Gravitational wave
signals entering the background histograms is commonly called
signal contamination of the $\rho - \xi^2$ background histograms. The contamination
caused by GW200129\_065458 in the background histogram of  bin 838 is shown in
\figref{fig:signal_contamination}. Since the GW signal gets added
to the background histogram, it occupies a region in $\rho - \xi^2$ space typical
of signals, but not of noise. As a result, we see a protrusion to the histogram, which
is generally how signal contamination manifests visually.

Signal contamination can result in the
$\rho - \xi^2$ histograms not accurately reflecting the noise characteristics
of the data, and as a result, the $\rho - \xi^2$ noise LR term will not
be calculated correctly. In general, it can cause the $\rho - \xi^2$ noise LR
term for GW candidates to be evaluated higher than it's true value,
leading to lower LR values of candidates. In short, signal contamination can
lower the sensitivity of the GW search.

\section{Removing contamination with the Background Filter}
\label{sec:removing_contamination}

To prevent any loss in sensitivity due to signal contamination, we need to
selectively remove the events in the $\rho - \xi^2$ background histograms
which originate from GW signals. The Background Filter is
a way to track the background in a time-dependent fashion so that we only use events
from times not corresponding to GW events to populate the 
background histograms. In this paper, we will describe the working of the
Background Filter when GstLAL is running in the low-latency online mode,
in which data is analyzed and results are produced in near real time~\cite{Ewing:2023}.

\subsection{Recording events}
The strategy of the Background Filter is to record the events
that are likely to have originated from GW signals,
and then after verification by the user, subtract them from the background
histograms. To associate events with a GW candidate, we need
to record the time at which they were found in the data, apart from their
$\vec{\rho}$ and $\vec{\xi^2}$ values. This increases the dimensionality of 
the parameters we need to store, and thus could potentially impact the
memory and storage used during
analysis. To prevent this, we record events only if they pass certain
constraints placed on their $\rho$, $\xi^2$, and time parameters.

\begin{figure*}
\centering
\begin{minipage}[b]{\columnwidth}
\includegraphics[width=\columnwidth]{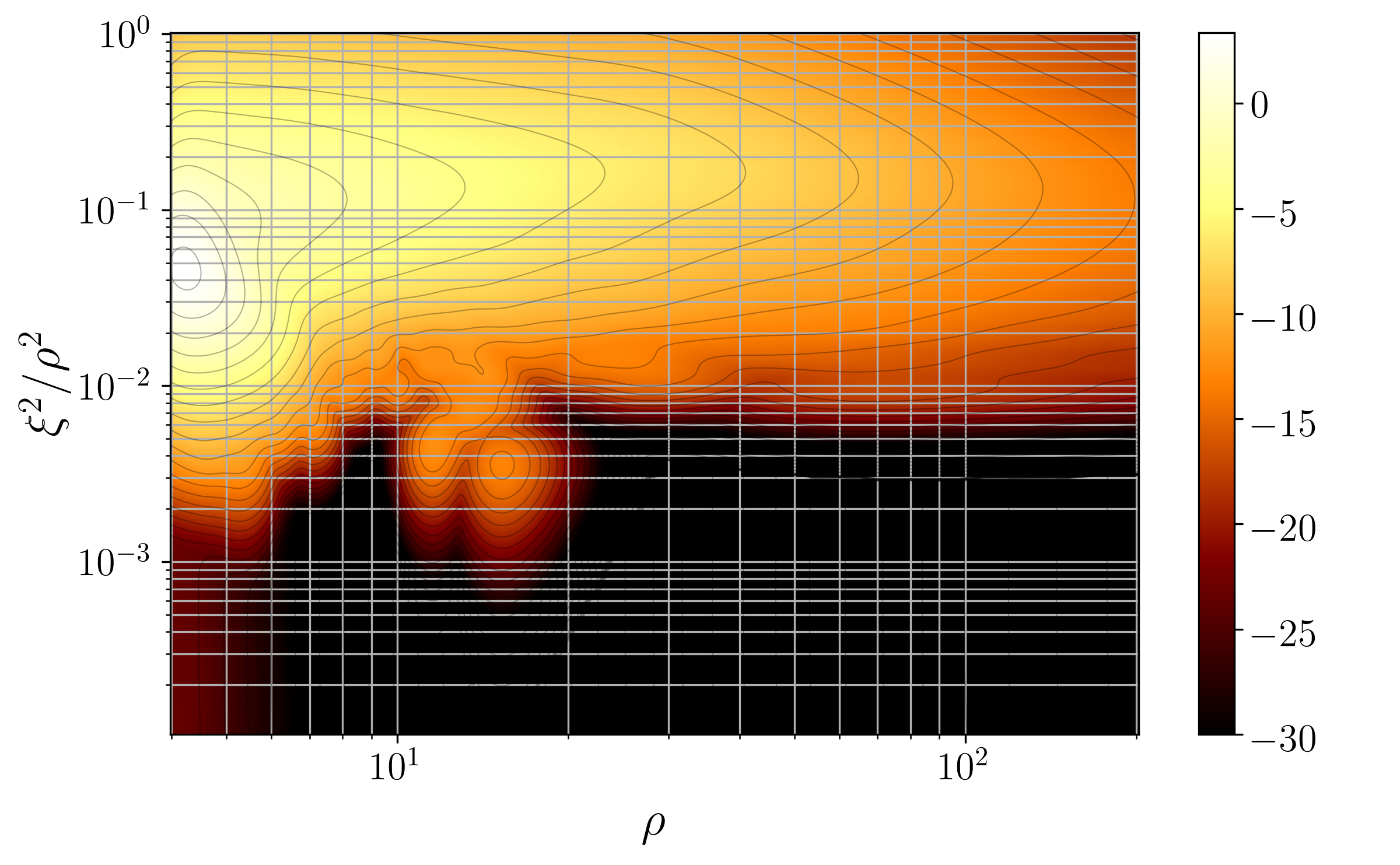}
\caption{\label{fig:signal_contamination}
An example of signal contamination in a $\rho - \xi^2$ histogram for Livingston.
The contamination can be seen as a protrusion to the histogram at
($\rho$, $\xi^2/\rho^2$) $\sim$ (15, 0.004), a region ususally occupied exclusively
by GW signals. This contamination was caused by GW200129\_065458
being recovered as a single event in this template bin, which is not the
best match bin for that GW candidate, as demonstrated in \figref{fig:svd}.
Note that kernel smoothing has been applied to this histogram.
}
\end{minipage}\hfill
\begin{minipage}[b]{\columnwidth}
\includegraphics[width=\columnwidth]{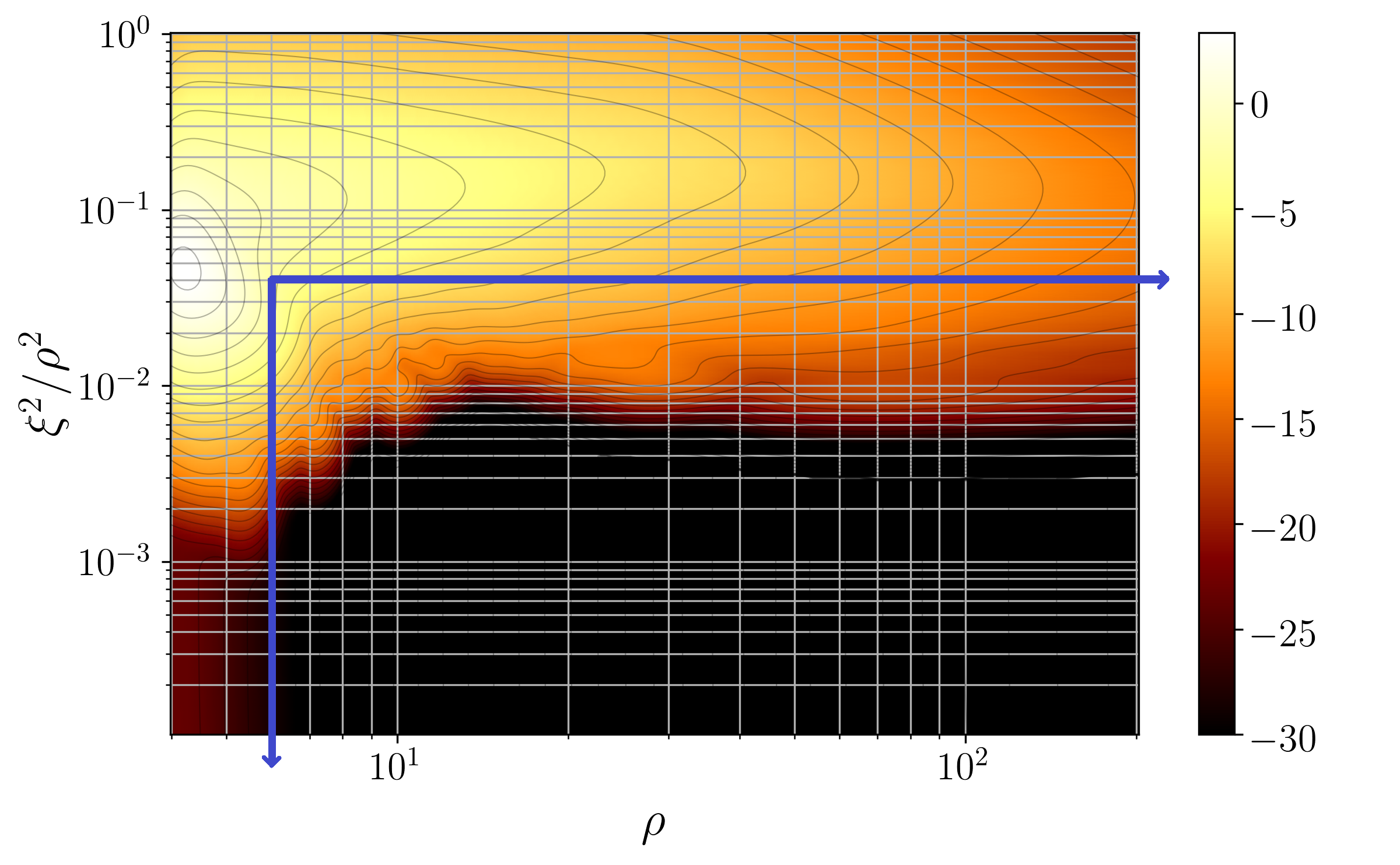}
\caption{\label{fig:clean}
The $\rho$ and $\xi^2$ constraints for recording events. The bottom
right area bounded by the blue lines is the area in which
the Background Filter records events. If the events also pass the time constraint,
the user can choose to remove them from the $\rho - \xi^2$ histogram.
The result of doing so, to remove the contamination
caused by GW200129\_065458 is also shown. The same histogram, without
using the Background Filter, and hence with contamination is shown
in \figref{fig:signal_contamination}. 
Note that kernel smoothing has been applied to this histogram.
}
\end{minipage}
\end{figure*}

The $\rho$ and $\xi^2$ constraints take the form of a bounding box in 
$\rho - \xi^2$ space, defined by $\rho > 6$ and $\xi^2/\rho^2 < 0.04$.
Qualitatively, $\xi^2$ represents how well the data fits the template, with
large values of $\xi^2$ meaning the data is dissimilar to the template.
Since in general, noise events will not fit the template well, they generally
have $\xi^2/\rho^2$ values that are greater than those of signals.
As a result, we only
expect events originating from GW signals to fall inside the bounding box.
This is shown in \figref{fig:svd}, where most of the high $\rho$ events
caused by GW200129\_065458 pass the $\rho$ and $\xi^2$ constraints, and are
recorded by the Background Filter.
The $\rho$ and $\xi^2$ constraints are shown on top of
a background histogram in \figref{fig:clean}

The time constraint makes use of the GstLAL online analysis' ability to
process data, generate events, assign LRs, and upload them to the Gravitational
Wave Candidate Event Database (GraceDB)~\cite{gracedb} in near real time.
A GW signal can create multiple contaminating events across
template bins. Only a small subset gets uploaded to GraceDB, since the events
are aggregated within some time window across bins before
uploading~\cite{Ewing:2023}, and the remaining
comtaminating events lie both before and after the uploaded events in time.
With this in mind, and in order to account for processing delays during a GstLAL
online analysis, the Background Filter keeps a temporary record of events passing
the $\rho$ and $\xi^2$ constraints, which occurred in the last 5000s. When an
event is uploaded to GraceDB, the events in the 10s window around it are found
from the temporary record of the last 5000s, and are then recorded
by the Bakground Filter permanently.

The threshold for uploading an event to GraceDB differs among different GstLAL
analyses, but it is often set to False Alarm Rate (FAR) $<$ 1 per hour. That means
all events recovered as a single event during coincident time, with $\rho > 6$,
$\xi^2/\rho^2 < 0.04$ and falling in a 10s interval around an event with
FAR $<$ 1 per hour are recorded by the Background Filter.

The $\rho$ and $\xi^2$ constraints, and the time constraint work together to
ensure that only events originating from GW signals are 
recorded by the Background Filter in most cases. As a result,
very few events are recorded by the Background Filter, 
in comparison to the number of events in the background histograms. This 
ensures that adding the Background Filter to a GstLAL analysis does not
affect its memory or disk usage significantly. The choice of these constraints,
and their impact on the performance of a GstLAL analysis are discussed in
\appref{app:constraints}.

\subsection{Removing contamination}

As explained in \secref{sec:signal_contamination}, since the $\rho - \xi^2$
noise LR term is calculated by evaluating the probability density function
represented by the background histrgrams at 
the relevant ($\vec{\rho}$, $\vec{\xi^2}$) value, we need the background
histograms to accurately reflect the detector noise characteristics for
that template bin. As much as possible, we need to take care not to let
events originating from signals enter the background histograms. In addition, we must
also make sure that events originating from noise are not removed from
the background histograms by the Background Filter. In most cases, the
$\rho$ and $\xi^2$ constraints along with the time constraint are
sufficient to ensure only events originating from signals are recorded
by the Background Filter.

However, in rare cases, such as when the GstLAL analysis uploads a 
false positive to GraceDB (also called a ``retraction"), these measures
might not be enough. Out of an abundance of caution, we leave the decision
of which events to remove from the background histograms to the user.
At any point during a GstLAL online analysis, the user can choose to inform
the analysis which events they are confident are GW
candidates. The message is communicated to the analysis in real time using
HTTP request methods, with the help of 
the Python \texttt{Bottle} module~\cite{bottle}. Then, out of all the
events that had been recorded by the Background Filter previously, it will
subtract those which fall within a 10s window of the given candidate, from
the background histograms. Thus, any contamination that that candidate
could have potentially caused is removed, and the LR of all future events
is evaluated using the modified $\rho$ - $\xi^2$ background histograms.

In \figref{fig:signal_contamination}, signal contamination caused by
GW200129\_065458 is shown. The same $\rho$ - $\xi^2$ histogram, but
with the Background Filter used to remove 
that contamination is shown in \figref{fig:clean}.

\section{Results}
\label{sec:results}

\subsection{Analysis methods}
To test the effect of signal contamination on the sensitivity of a GstLAL
analysis, and the ability of the Background Filter to remove the
contamination, we analyze a week of O3 data~\cite{gwosc}, from Apr 18 2019 16:46 UTC
to Apr 26 2019 17:14 UTC, in three different ways. First, we perform
a control run without any GW signals. Next, to simulate the effect of GW
signals, we add ``blind injections". The concept of blind injections, 
and the set of blind injections that we used are explained in the following subsection. Finally, we
perform a ``rerank" with the Background Filter enabled, in which LRs
are recomputed and significance assignment is done again, but the 
filtering stage of the GstLAL analysis is taken from the blind injection
analysis, since the $\rho$ and $\xi^2$ values of analyzed
events are not affected by the Background Filter, only the LRs and the
False Alram Rates (FARs) are. Hence, the rerank corresponds to the case
with blind injections present, and the Background Filter being used.

An important point to note is that since we are using blind injections
to replicate the effect of GW signals, we know with
certainty the times when GW signals will occur. This 
allows us to use the Background Filter on all those times, thus
presenting a best-case scenario for the efficacy of the Background Filter.

This chunk of data contains two known GW candidates reported in
GWTC-2.1~\cite{gwtc-2.1} and elsewhere~\cite{Abbott:2020uma, LIGOScientific:2021djp},
GW190421\_213856 and GW190425. However, since we use the 
Background Filter only on the times of the blind injections, any contamination
and subsequent loss in sensitivity caused by either of the two candidates will
be present in all three analyses that we perform, and will not affect the 
evaluation of the performance of the Background Filter.

\subsection{Simulation Set}
\begin{figure}
\includegraphics[width=\columnwidth]{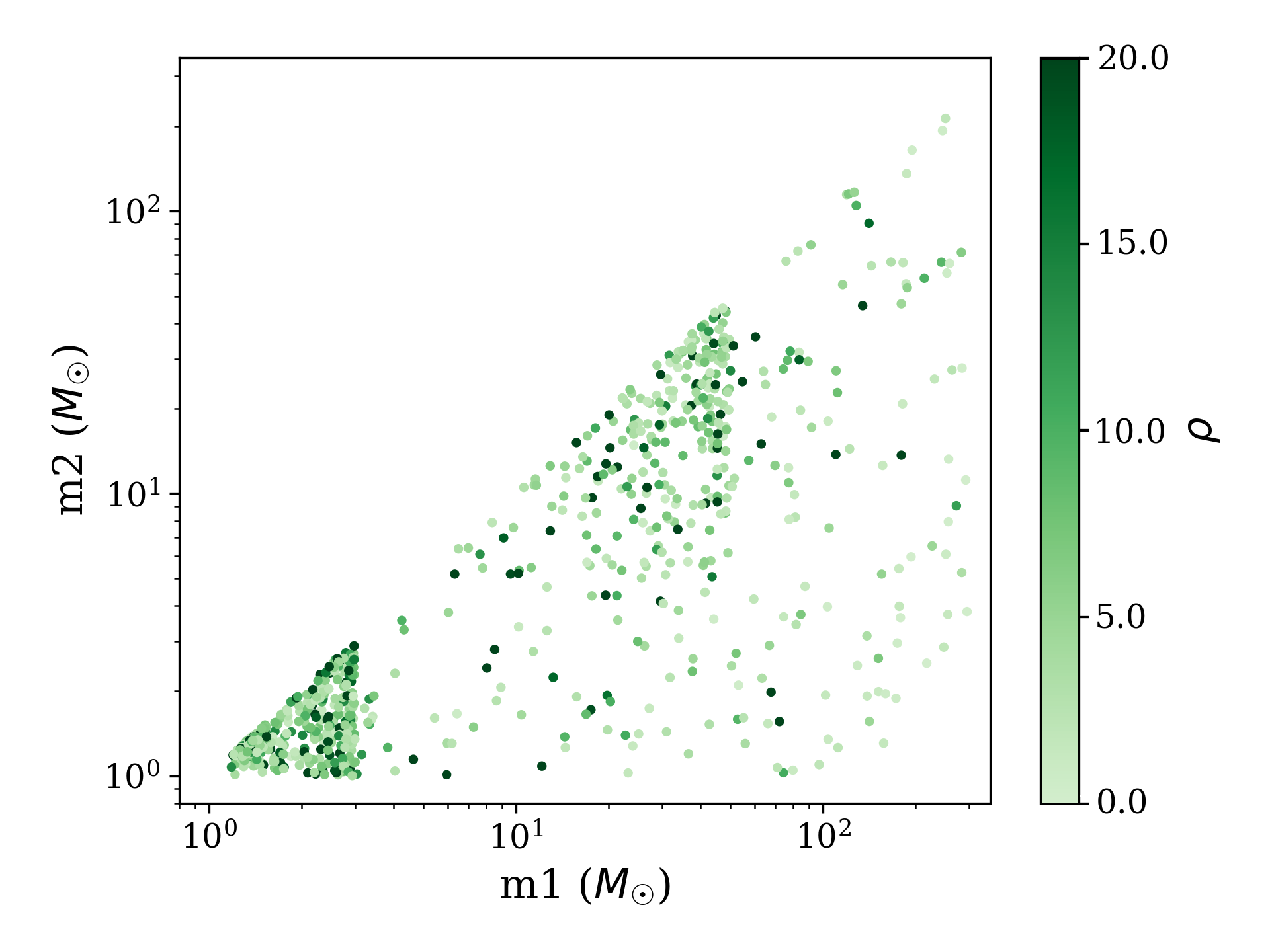}
\caption{\label{fig:blind_inj}
The distribution of component masses of the blind injection set,
colored by injected $\rho$. Blind
injections are used to replicate the contamination caused by GW
signals in the data.
}
\end{figure}

Blind injections are simulated GW signals that are added
to the data which we analyze and collect background events from (in
contrast to regular injections, from which we do not collect background
events). We use a set of 868 blind injections distributed across the 
binary black hole (BBH), binary neutron star (BNS), neutron star-black
hole (NSBH), and intermediate-mass
black hole (IMBH) parameter spaces. The blind injection set comprises of
three subsets, a BNS subset, a BBH subset and a broad subset, with the
BNS subset containing half of the total blind injections, and the BBH
and broad subsets containing a quarter each.
The BNS subset has component
masses distributed uniformly from 1 to 3 $M_\odot$, and the z-components
of dimensionless spin (which are parallel to the orbital angular momentum
of the binary) distributed uniformly from -0.05 to 0.05.
The BBH subset has component masses distributed uniformly from 5 to 50
$M_\odot$ and the z-components
of dimensionless spin distributed uniformly from -0.99 to 0.99.
The broad subset spans all four parameter spaces mentioned above. It is
distributed uniformly in the log of the component masses
from 1 to 300 $M_\odot$ and has the z-components
of dimensionless spin distributed uniformly from -0.99 to 0.99.
In addition to the definitions of the BNS and BBH parameter spaces
provided above, and the implied NSBH parameter space definitiion,
for the purpose of this paper, we shall consider the parameter space
with either component mass greater than 50 $M_\odot$ to be the IMBH space.
The distribution of the blind injection set in the two component
masses can be seen in \figref{fig:blind_inj}.

A point to note is that even though 868 blind injections
may sound high, most of these are too quiet to be recovered, as shown in
\figref{fig:blind_inj}, and hence won't cause any contamination. The
result of the analysis shows that only $\sim$ 200 blind injections
are loud enough to be recovered. We will also see later that BNS and NSBH template
bins are not affected by signal contamination to a significant degree.
As a result, only the BBH and IMBH injections will contribute
to contaminating the background. Given the high number of GW candidate
events we expect to detect in O4, this is a reasonable 
representation of the total amount of signal contamination we expect to see.

We also perform an injection
campaign to calculate the sensitivity of the analysis, both with
and without the application of the Background Filter. The injection set
is distributed similarly to the blind injection set, but with
a total size of 86,606 injections. It is important to note that the injections and 
blind injections are analyzed separately, with the blind injections
affecting injection recovery only through the background events they
add to the $\rho$ - $\xi^2$ background histograms.

\subsection{Sensitivity Improvements}
In order to estimate the sensitivity of a search, we use the sensitive
volume-time (\textit{VT}) as a measure. The volume that we analyze is determined
by the efficiency of recovering injections at a given FAR and redshift,
and T is the livetime of the analysis. We calculate \textit{VT}
separately for injections falling in four different chirp mass bins. The
first is from 0.5 to 2 $M_\odot$, the second from 2 to 4.5 $M_\odot$, the
third from 4.5 to 45 $M_\odot$, and the final one is from 45 to 450
$M_\odot$. The reason for calculating \textit{VT} separately for different mass
bins is so that we have an idea about how sensitive the analysis is 
for different source categories, with the four mass bins roughly
corresponding to BNS, NSBH, BBH and IMBH
source categories respectively.

\begin{figure}
\includegraphics[width=\columnwidth]{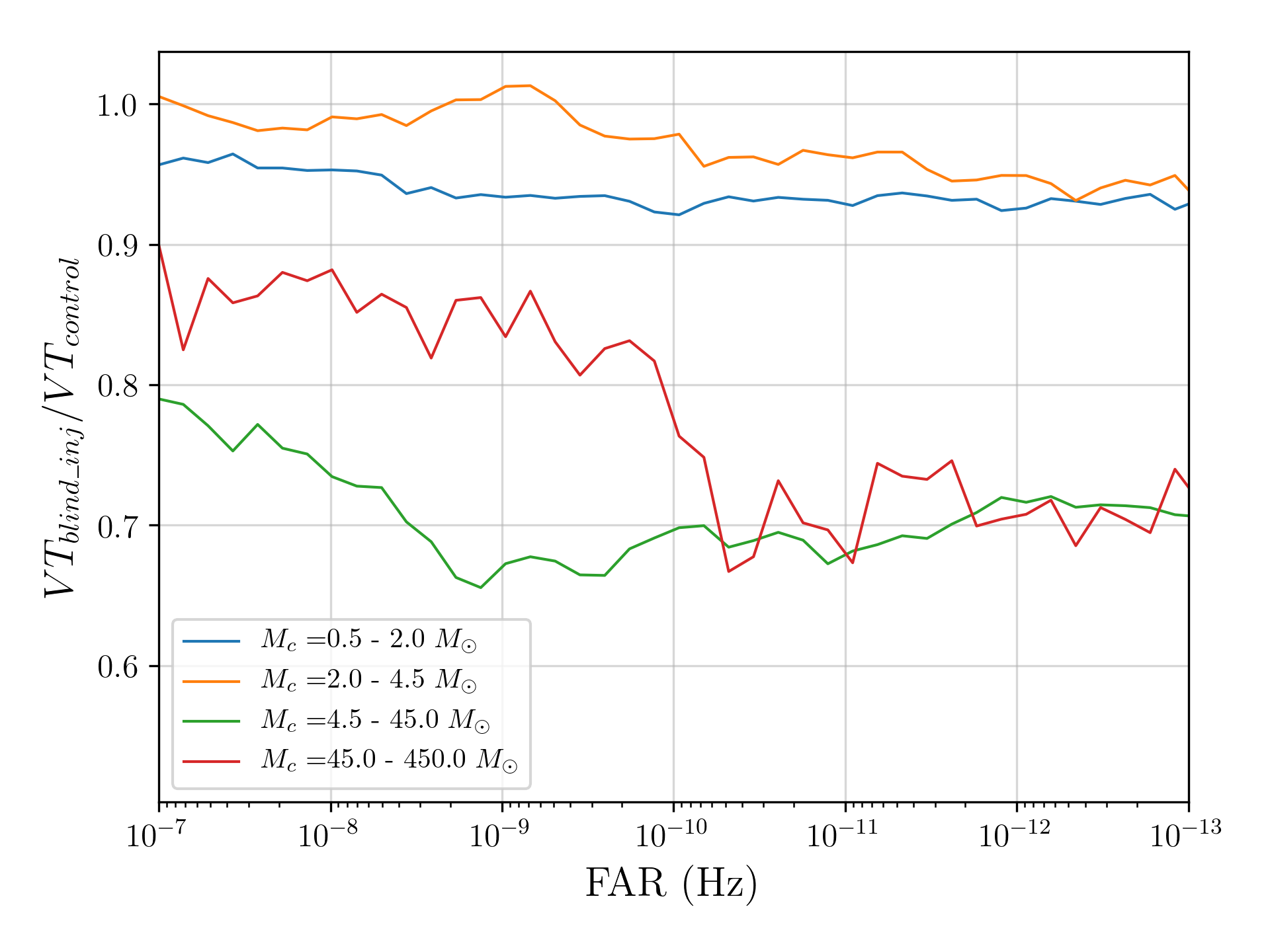}
\caption{\label{fig:vt_01}
The decrease in \textit{VT} caused by signal contamination due to the 
presence of blind injections in the data. The two highest mass
bins are the most affected. The presence of GW signals
will also have a similar effect.
}
\end{figure}

Comparing the blind injection analysis with the control run,
signal contamination due to the presence of blind injections causes
a small ($\sim$ 5\%) decrease in \textit{VT} in the two lowest mass bins, but
causes a significant ($\sim$ 20-30\%) decrease in \textit{VT} in the two highest
mass bins. This is shown in \figref{fig:vt_01}. High mass templates have a
greater match with their neighbouring templates, and with themselves across time,
as compared to low mass templates. We hypothesize that this causes a single high
mass GW signal to be recovered multiple times across template bins and time with
suboptimal $\rho$, leading to more signal contamination in the high mass template
bins than in the low mass ones. This is discussed in more detail in \appref{app:bins}.

\begin{figure}
\includegraphics[width=\columnwidth]{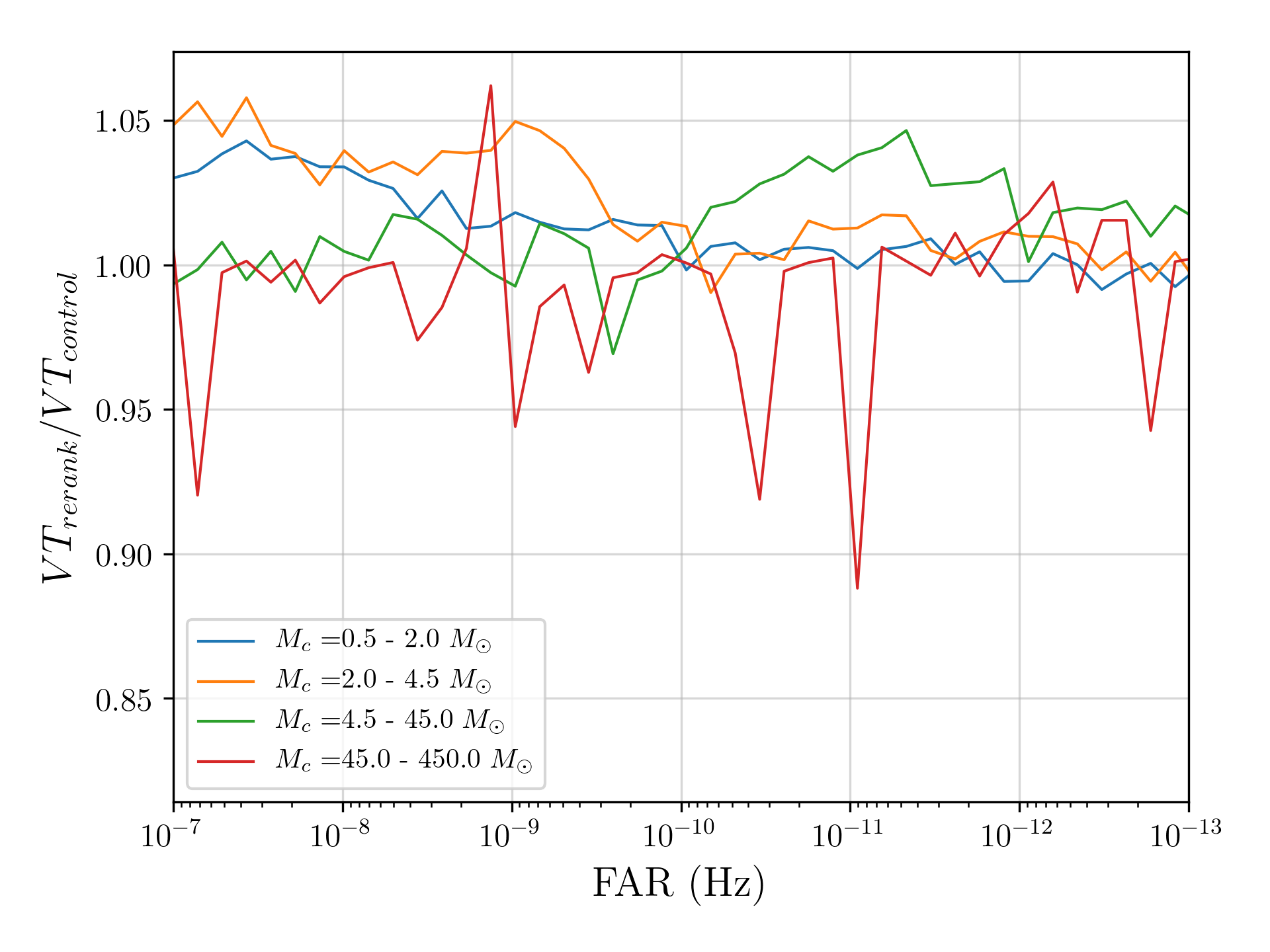}
\caption{\label{fig:vt_04}
The \textit{VT} of the rerank, which has blind injections with the Background
Filter applied, compared to that of the control run, which has
neither. The fact that all four lines are close to 1 tells us
that the Background Filter is successful in removing close to
all of the contamination caused by the presence of the blind
injections in the data. The peaks and dips in the highest mass bin
curve are explained by the smaller number of injections in this bin
as compared to other bins, leading to greater variance.
}
\end{figure}

Next, to check the efficacy of the Background Filter in removing
contamination, we compare the \textit{VT} of the rerank to the \textit{VT} of the
control run. Despite the
presence of blind injections in the data, The Background Filter
mitigates the effect they have on the background histograms, and
sensitivities of all four mass bins are essentially the same as what
they were in the control run.
This is shown in \figref{fig:vt_04}. This represents a 20-40\% increase
in the sensitivities of the two high mass bins, in the case of the
rerank, as compared to that of the blind injection analysis. We can 
conclude that in the best case, the Background Filter is 
successful in removing all the contamination that the blind
injections cause. Since the number of blind injections we
used was a high estimate of the number of GW events
we expect to see in O4, this means that by using the Background
Filter, we do not expect signal contamination to be a significant
problem during O4. To test our readiness for O4, GstLAL has
participated in a mock data challenge, where an online analysis is
run over forty days of O3 data~\cite{Ewing:2023}. The Background
Filter was deployed in this analysis, and it was able to remove
all instances of signal contamination.

\section{Conclusion}
GstLAL constructs $\rho$ - $\xi^2$ background histograms to calculate
the $P\left(\vec{\rho}, \vec{\xi^2} \mid t_\mathrm{ref}, \bar{\theta}, \mathcal{H}_\mathrm{n}\right)$
term in the likelihood ratio. However, GW
signals in the data can cause the background histograms to
be incorrectly constructed. This is called signal contamination,
and it leads to the sensitivity of the GstLAL analysis being 
lowered.

The Background Filter is a novel way to remove the contamination.
It records the events that populate the background
histograms which satisfy two constraints. The first is that the
event must fall in an area in $\rho$ - $\xi^2$ space
consistent with GW signals. The second is that it
must fall in a 10s window around a significant event. The user
then identifies which of the significant events originate from
GW signals. The user communicates this to the
GstLAL analysis in real time, and then the events recorded by the
Background Filter corresponding to the times identified by the user
are subtracted from the background histograms. Thus, signal
contamination is removed from the background histograms.

To test the efficacy of the Background Filter, we ran a GstLAL
analysis over a week of O3 data, with simulated gravitational-wave
signals injected into the data. We found that signal
contamination primarily affects the high mass bins. The 
sensitivity of these bins decreased by 20-30\% due to the
presence of the gravitational-wave signals. By applying the 
Background Filter, we were able to increase the sensitivity back
to what it was without the injected gravitational-wave signals.
This shows that the Background Filter is effective in removing 
close to all the signal contamination.
With a high rate of gravitational-wave events expected during
O4, the Background Filter will be an important tool in
improving the sensitivity of the GstLAL analysis.

\begin{acknowledgments}
The authors are grateful for computational resources provided by the LIGO
Laboratory and supported by National Science Foundation Grants PHY-0757058
and PHY-0823459.  This material is based upon work supported by NSF's LIGO
Laboratory which is a major facility fully funded by the National Science
Foundation.  LIGO was constructed by the California Institute of Technology
and Massachusetts Institute of Technology with funding from the National
Science Foundation (NSF) and operates under cooperative agreement
PHY-1764464.  The authors are grateful for computational resources provided
by the Pennsylvania State University's Institute for Computational and Data
Sciences (ICDS) and by the California Institute of Technology, and support
by NSF PHY-\(2011865\), NSF OAC-\(2103662\), NSF PHY-\(1626190\), NSF
PHY-\(1700765\), NSF PHY-\(2207728\), and NSF PHY-\(2207594\).

This research has made use of data or software obtained from the Gravitational Wave Open Science Center (gwosc.org), a service of LIGO Laboratory, the LIGO Scientific Collaboration, the Virgo Collaboration, and KAGRA. LIGO Laboratory and Advanced LIGO are funded by the United States National Science Foundation (NSF) as well as the Science and Technology Facilities Council (STFC) of the United Kingdom, the Max-Planck-Society (MPS), and the State of Niedersachsen/Germany for support of the construction of Advanced LIGO and construction and operation of the GEO600 detector. Additional support for Advanced LIGO was provided by the Australian Research Council. Virgo is funded, through the European Gravitational Observatory (EGO), by the French Centre National de Recherche Scientifique (CNRS), the Italian Istituto Nazionale di Fisica Nucleare (INFN) and the Dutch Nikhef, with contributions by institutions from Belgium, Germany, Greece, Hungary, Ireland, Japan, Monaco, Poland, Portugal, Spain. KAGRA is supported by Ministry of Education, Culture, Sports, Science and Technology (MEXT), Japan Society for the Promotion of Science (JSPS) in Japan; National Research Foundation (NRF) and Ministry of Science and ICT (MSIT) in Korea; Academia Sinica (AS) and National Science and Technology Council (NSTC) in Taiwan.
\end{acknowledgments}

\appendix

\section{Choice of constraints, and their impact on performance}
\label{app:constraints}
With the constraints described in \secref{sec:removing_contamination},
the Background Filter does not consume too many resources. When looking at
a month-long GstLAL analysis, we found that on average,
it adds $\sim$ bytes to kilobytes to the data products stored by a 
GstLAL analysis for every template bin. We didn't see any significant
increase to the memory used by the GstLAL analysis either. \figref{fig:vt_04}
shows that with these constraints, the Background Filter is effective in
removing close to all contamination.

\begin{figure}
\includegraphics[width=\columnwidth]{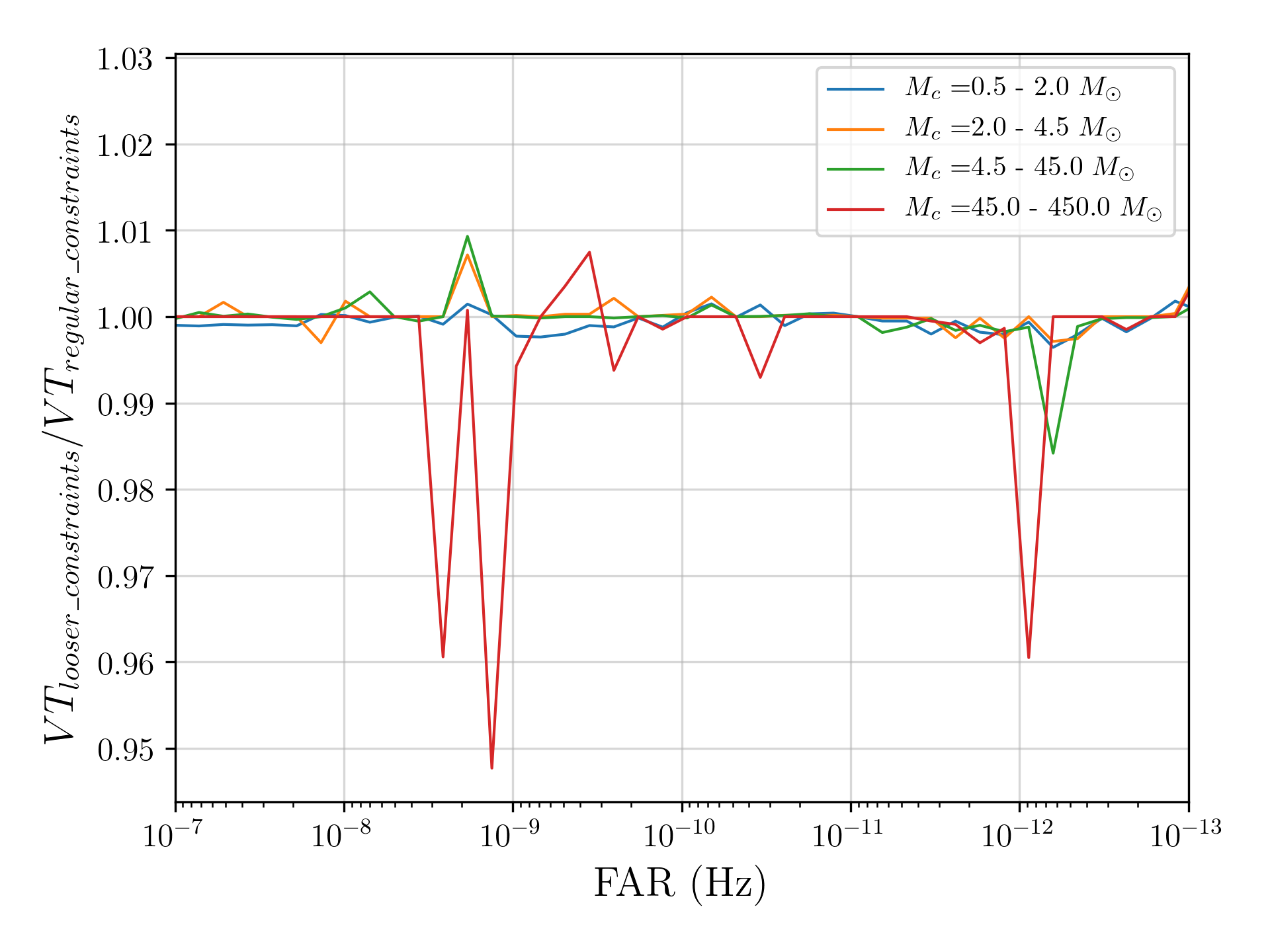}
\caption{\label{fig:vt_35}
The \textit{VT} with the Background Filter recording events with looser constraints,
as compared to the \textit{VT} with the Background Filter recording events with the
regular constraints. The fact that the \textit{VT} ratios for all four mass bins are
close to 1 shows that loosening the constraints does not improve sensitivity.
The peaks and dips in the highest mass bin
curve are explained by the smaller number of injections in this bin
as compared to other bins, leading to greater variance.
Both analyses included the 868 blind injections described in \secref{sec:results}.
The regular constraints are described in \secref{sec:removing_contamination},
whereas the looser constraints are described in \appref{app:constraints}.
}
\end{figure}

To check if there is any improvement to the sensitivity upon loosening
the $\rho$ and $\xi^2$ constraints, we performed the same analysis as 
described in \secref{sec:results}, but with the $\rho$ and $\xi^2$ constraints
changed to record events with $\rho > 6$ and $\xi^2/\rho^2 < 0.4$. This
broader bounding box for recording events did not have any noticeable
effect on the sensitivity. This is shown in \figref{fig:vt_35}. However, loosening
the constraints did increase memory usage of the GstLAL analysis to a noticeable degree.

We do not expect that loosening the $\rho > 6$ constraint or the time constraint
would increase sensitivity, since the extra events collected by changing these
constraints would be no more significant than noise. This discussion, along
with \figref{fig:vt_04} show us that the existing constraints used by the
Background Filter satisfy all our requirements.

\section{Differing impacts of singal contamination of the sensitivities of template bins}
\label{app:bins}
As discussed in \secref{sec:results}, signal contamination only causes a 5\% decrease
in the \textit{VT} of low mass template bins, such as the BNS and NSBH bins, whereas it
causes a 20-30\% decrease in the \textit{VT} of high mass template bins, such as the BBH
and IMBH bins. This is despite the fact that there are more blind injections in the low
mass parameter spaces than in the high mass ones. We conjecture two reasons for this,
the first relating to how the correlation among neighbouring templates changes
with mass, and the second relating to how the correlation of a templates with itself across
time changes with mass. For the remainder of this section, we shall treat BNS
template bins as representative of all low mass bins, and IMBH template bins as 
representative of all high mass bins.

\begin{figure*}
\includegraphics[width=\textwidth]{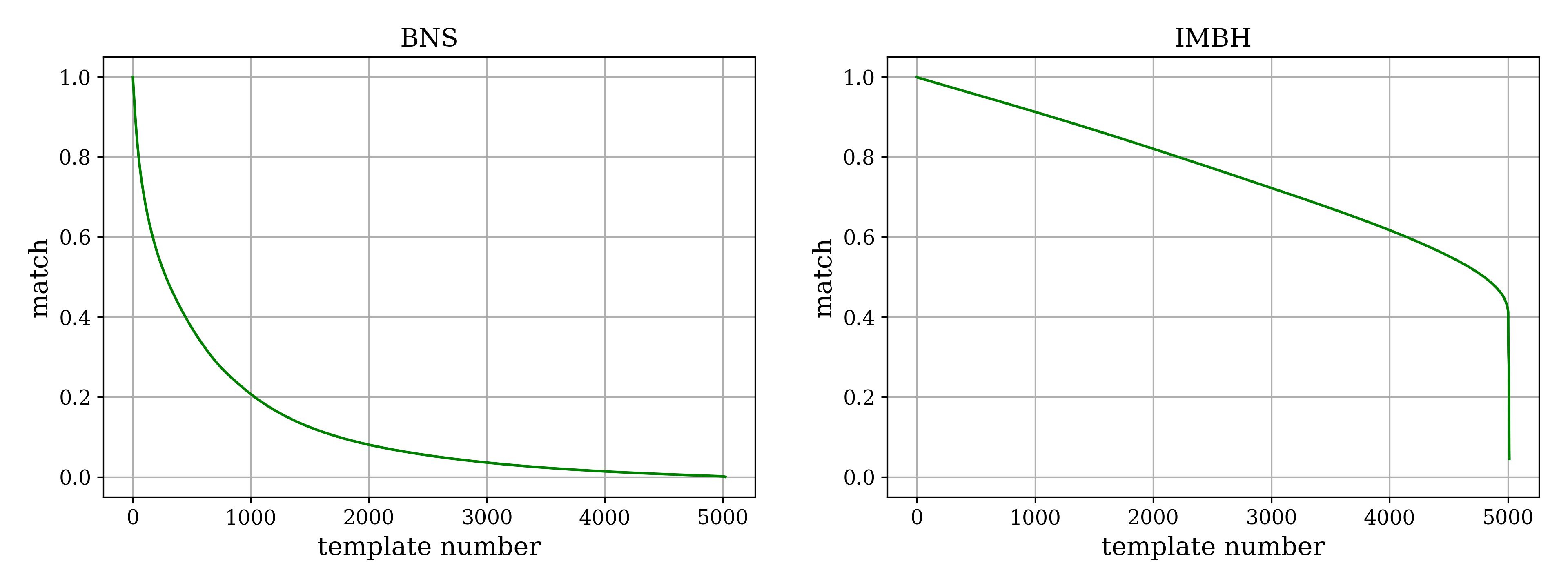}
\caption{\label{fig:bank_correlation}
The average bank correlation function of a BNS template in descending order of template
match, as compared to that of an IMBH template, calculated for the five closest template
bins. Since IMBH templates correlate well with other IMBH templates across template bins,
an IMBH GW signal will be recovered by multiple template bins, increasing the probability
of signal contamination. This is not the case for BNS template bins, and it is more likely a BNS
GW signal will be recovered by only one template bin, resulting in fewer cases of 
signal contamination.
}
\end{figure*}

The ``bank correlation function" of a template measures how well it
matches with other templates in the template bank. This calculation is similar to
how $\rho$ is calculated, except that the match is calcualted between two templates with
no time shift between them. To see how the bank correlation function of templates changes
with mass, we took 5 BNS template bins (corresponding to $\sim$ 5000 templates),
calculated the bank correlation of every combination of templates, and plotted the average
bank correlation function in descending order of template match. We then did the same for
5 IMBH template bins. The results are shown in \figref{fig:bank_correlation}. The fact that the BNS
bank correlation function drops sharply as compared to the IMBH one, means that there
are many IMBH templates, across template bins that can recover a given IMBH GW signal with a 
lower $\rho$ than the best template, but only a few BNS templates that can recover a given BNS GW signal.
This means a high mass GW signal will create many events, increasing the probability of 
signal contamination. This is not a problem for the GW candidates reported by GstLAL, since
``event clustering"~\cite{Messick:2016aqy} ensures that only the best candidate in an 8s
window survives.

\begin{figure*}
\includegraphics[width=\textwidth]{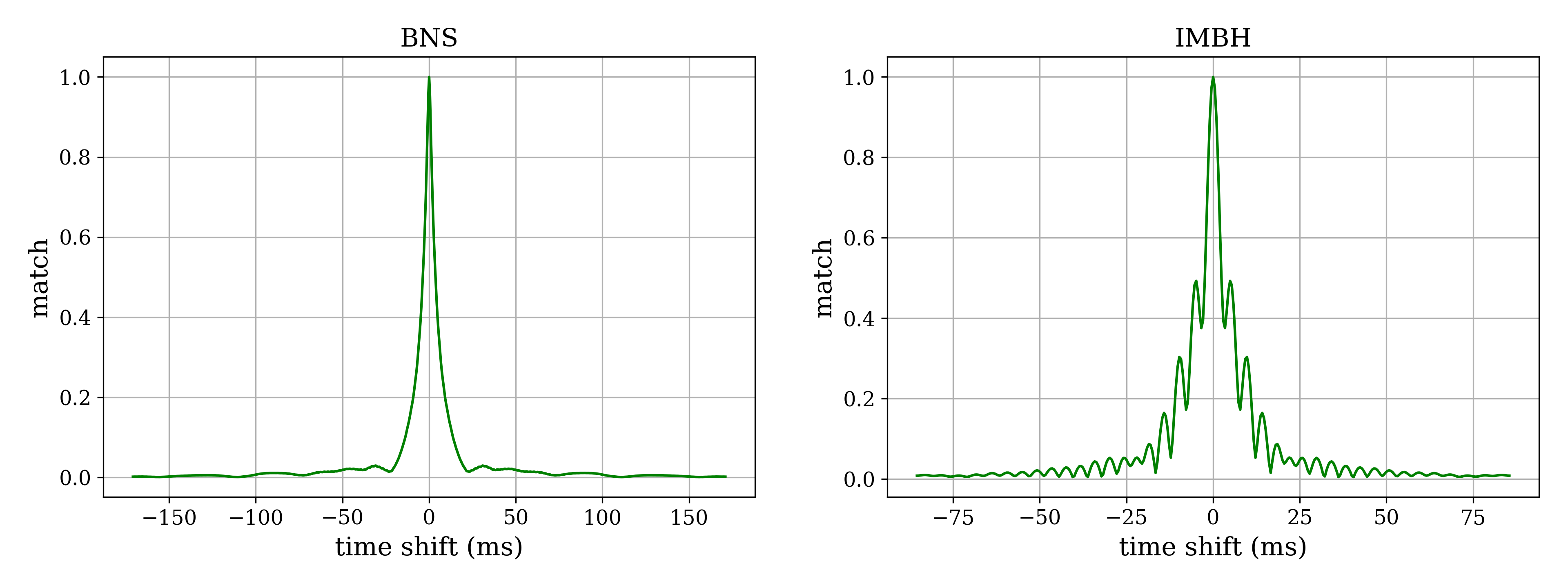}
\caption{\label{fig:auto_correlation}
The autocorrelation function of a BNS template, as compared to that of an IMBH template. Since
there are multiple peaks in autocorrelation function of the IMBH template, a quiet IMBH GW signal
could be recovered in different detectors at different times, corresponding to the different
peaks in the IMBH autocorrelation function. This will result in the GW singal being recovered
as multiple single detector events rather than a single coincident event, which leads to
signal contamination. Since the BNS autocorrelation function does not have multiple peaks,
signal contamination is less likely for BNS template bins.
}
\end{figure*}

The ``autocorrelation function" of a template measures how well it matches with a time-shifted
version of itself, similar to how $\rho$ calculates the match between the data and a 
time-shifted template. The autocorrelation function of a typical BNS template and a typical
IMBH template are shown in \figref{fig:auto_correlation}. The IMBH autocorrelation function 
has multiple secondary peaks $\sim$ 5-10ms away from the primary one. We conjecture that
in the case of an IMBH GW singal with low $\rho$ or in high noise, an IMBH template could recover
the signal in different detectors at different times, corresponding to the different peaks
in the IMBH autocorrelation function. This would cause the signal to be recovered as multiple
single detector events, instead of a single coincident one, leading to signal contamination
of the high mass template bins. Again, this is not a problem for the GW candidates reported by
GstLAL, due to event clustering. Since all the secondary peaks in the autocorrelation function of
an IMBH template lie well within an 8s window, multiple single detector events will be clusterd out,
and only the best one will survive.

For high mass bins, the bank correlation factor increases the probability of low $\rho$ events
getting created by a GW signal, and the autocorrelation factor increases the probability of signal
contamination from those events. These two factors compound each other's effect,
and as a result, we see a much higher
impact of signal contamination in the high mass template bins than in the low mass ones.

\bibliography{references}

\end{document}

%% file: author_list.tex
\author{Prathamesh Joshi \orcidlink{0000-0002-4148-4932}}
\email{prathamesh.joshi@ligo.org}
\affiliation{Department of Physics, The Pennsylvania State University, University Park, PA 16802, USA}
\affiliation{Institute for Gravitation and the Cosmos, The Pennsylvania State University, University Park, PA 16802, USA}

\author{Leo Tsukada \orcidlink{0000-0003-0596-5648}}
\affiliation{Department of Physics, The Pennsylvania State University, University Park, PA 16802, USA}
\affiliation{Institute for Gravitation and the Cosmos, The Pennsylvania State University, University Park, PA 16802, USA}

\author{Chad Hanna}
\affiliation{Department of Physics, The Pennsylvania State University, University Park, PA 16802, USA}
\affiliation{Institute for Gravitation and the Cosmos, The Pennsylvania State University, University Park, PA 16802, USA}
\affiliation{Department of Astronomy and Astrophysics, The Pennsylvania State University, University Park, PA 16802, USA}
\affiliation{Institute for Computational and Data Sciences, The Pennsylvania State University, University Park, PA 16802, USA}